# Atomically Resolved Spin-Dependent Tunnelling on the Oxygen-Terminated Fe3O4 (111)

N. Berdunov, S. Murphy, G. Mariotto, and I. V. Shvets

SFI Nanoscience Laboratory, Physics Department, Trinity College, Dublin 2, Ireland

We employ spin-polarized (SP) STM to study the spin-dependent tunneling between a magnetite (111) sample and an antiferromagnetic tip through a vacuum barrier at room temperature. Atomic scale STM images show significant magnetic contrast corresponding to variations in the local surface states induced by oxygen vacancies. The estimated variations in tunneling magnetoresistance (TMR) of 250% suggest that the spin-transport properties are significantly altered locally by the presence of surface defects.

## Introduction

Magnetite, $Fe_3O_4$, predicted to be a half-metallic ferromagnet, attracts a lot of interest from the spin electronics community. It is expected that a magnetic tunnel junction (MTJ) with a $Fe_3O_4$ electrode could exhibit a high TMR effect, but the reported values are much lower than expected[1,2]. However, photoelectron spectroscopy measurements on the magnetite surface do show much greater values of the Spin Polarization (SP)[3]. It was suggested that the reduction of SP is due to the disorder at the electrode/barrier drastically changing the spin-polarized properties of the interface[4,5,6,7]. In complex structures, like magnetite, the surface can present a number of non-identical terminations, which possess a variety of spin-electronic properties. For example, it was shown that an oxygen layer deposited on a magnetic electrode could change MTJ properties drastically, even reversing the sign of spin-polarization[8].

In the present work we apply Spin Polarized Scanning Tunneling Microscopy (SP-STM) measurements to analyze the electronic structure of the oxygen-terminated magnetite (111) surface on the atomic scale. We aim to understand the impact of point defects on the surface electronic structure and on the spin dependent tunneling between a magnetite sample and an antiferromagnetic MnNi tip through a vacuum barrier. The potential of SP-STM for studies of surface magnetic properties has been demonstrated in recent experimental and theoretical works[5,9,10]. Spin-dependent tunneling can be achieved by selecting the tunneling conditions for the majority and minority spin states. Such kind of selective SP-STM techniques can be broadly divided into two groups. The first one utilizes a ferromagnetic tip with the magnetization switched by a local magnetic field[11,12]. However, there are no reports of atomic resolution achieved with this approach so far. The second approach uses an antiferromagnetic tip to scan the surface, where the applied magnetic field applied is strong enough to alter the sample magnetization. In an earlier theoretical work it has been shown that spin-contrast can be achieved in the case of tunneling between antiferromagnet and ferromagnet electrodes[13]. Atomically resolved SP-STM data employing Cr[14] and MnNi[15,16] tips have been recently reported. It is worth pointing out that SP-STM does not provide the absolute value of spin polarization, as there are uncertainties in the surface/tip state contributions to tunneling. Nonetheless, by analyzing SP-STM results, in particular, with the help of *first principles calculations*, further insights into the local density of states on the surface and their contribution to the tunnel current can be obtained [9].

The experiments have been performed in ultra-high vacuum (UHV) at room temperature. In our SP-STM measurements, tips made of antiferromagnetic MnNi alloy were employed[17]. A synthetic magnetite (111) single crystal has been used in these experiments. The crystal structure of magnetite allows for six ideal bulk terminations in the (111) plane (Fig.1). In the present Letter we limit our discussion by considering the case of the two oxygen terminations. Both present close-packed oxygen layers with identical atomic periodicities. One appears as an oxygen monolayer on top of an octahedral Fe layer, while the other covers a multilayer of tetrahedral and octahedral Fe atoms. By choosing either a reducing or an oxidizing sample preparation procedure, it is possible to achieve two stable terminations. One is referred as the Fe-tetrahedral termination[18], while the other represents an oxygen termination with a long-range order on the surface[19]. The latter is a subject of this study.

*STM results*

The clean magnetite sample was annealed in an oxygen atmosphere of $10^{-6}$ mbar at 950K for 15 min followed by cooling in the oxygen atmosphere. This sample preparation procedure leads to the formation of a well-defined hexagonal superlattice with a periodicity of 42±3 Å. The superstructure is highly regular and covers almost the entire sample surface. The high-resolution STM image in Fig.2(a) shows the atomic arrangement within the superstructure. One can see that the superstructure consists of three distinct areas, marked as areas I, II and III. Detailed analysis shows that area I has an average periodicity of 3.1±0.1 Å, while areas II and III have a periodicity of 2.8 ± 0.1 Å, which is consistent with the low-energy electron diffraction (LEED) pattern[19]. As we have shown in an earlier publication[19], the superstructure represents an oxygen-terminated magnetite bulk, which reconstructs due to an electron-lattice instability, like a polaron or a charge density wave. Thus, the STM image in Fig.2(a) comprises a lattice of oxygen sites on the top of an iron layer. A number of point defects seen in Fig.2(a) correspond to missing oxygen atoms. We have further shown that for a critical density of oxygen vacancies of some 30%, the superstructure disappears[20]. This means that the oxygen topmost layer is responsible for the electronic transformation in the sub-surface layer, and for the changes in the Fe-states leading to the superstructure formation.

In our SP-STM experiments a magnetic field of 60 mT was applied parallel to the surface during the STM scan. The STM images were taken on the same sample within the space of a few hours between the non-magnetic and magnetic experiments, no additional sample treatment between the experiments were performed. The spin-contrast achieved in STM images when the magnetic field was switched off/on is demonstrated in Fig.2(a) and (b) respectively. It is clear that the appearance of the superstructure and the corrugation difference between different areas is almost unaffected by the magnetic field. However, major changes occur on the atomic scale in proximity of the oxygen vacancies. Three bright spots appear in the vicinity of the defects as can be seen in Fig.2(b). The 6 Å separation between the spots and their positions correspond to those of Fe ions in the layer underneath the topmost oxygen lattice. The effect becomes more pronounced when the density of defects rises. Fig.2c shows the case when each defect comprises a few missing oxygen atoms i.e. vacancy cluster. The bright spots appear on the periphery of the extended defects and still have the spacing and symmetry corresponding to the Fe atoms.

We can quantify the observed spin polarized effect in terms of the tunneling conductance variation around the surface defects

$$\sigma = (G_{60mT} - G_0)/G_0 , \qquad (1)$$

where $G_{60mT}$, $G_0$ are the tunneling conductances with and without the applied magnetic field. To estimate the tunneling conductances, we can use a simplified approach where the tunnel current $I_t$ depends on the bias voltage $U$, the distance $d$, the energy of the barrier between the two electrodes $\varphi$ and constant $A = 1.025$

$$I_t \sim (U/d_i) \exp(-A \, \varphi_i^{-1/2} \, d_i)$$

Where index i=1,2 corresponds to the case when the tip is positioned above an oxygen atom and above a Fe atom, respectively (see Fig.2d). In our experiment, the tunnel current is $I_t$=0.1nA, and the bias voltage is $U$=1V. The tip displacement over the point of the corrugation maxima in the STM image is equal to 0.3 Å. Assuming that the tip-surface distance is 5 Å, we can calculate the change in the barrier energy and the corresponding tunneling conductance. Their substitution into formula (1) yields a $\sigma$ value of 2.5, which indicates a significant spin polarization effect. Our assumption regarding the tip-sample distance is based on the conclusion of the paper [21, 22]. In fact, the broad distance range 4-10 Å will still give us a high value of $\sigma$ from 3.8 to 0.9.

The formula (1) we used is similar to *Julliere's* definition of TMR [23]:

TMR=$(G_p$-$G_{ap})/G_{ap}$ * 100%. Where $G$p, $G$ap are the tunneling conductances in case of parallel and antiparallel magnetizations at the electrodes. Assuming $G_{60mT} = G_p$; $G_0 = G_{ap}$, this corresponds to a 250% TMR effect observed. Although, in the absence of a magnetic field it is not necessarily true that we fulfill the conditions for the minimum conductance, so that the correct correspondence is $G_0 \geq G_{ap}$. Therefore the calculated TMR value should be taken as the lower limit.

*First Principles Calculations* and Discussion

To analyze the SP-properties of the oxygen-terminated magnetite (111) surface we have performed DFT calculations for the ideal oxygen-terminated magnetite (111) vacuum slab (vacuum/magnetite/vacuum interface) with and without the presence of a surface defect. The CASTEP program [24] as a module of Materials Studio was used in our calculations. A one unit cell vacuum slab similar to that shown in Fig.1 has been constructed. The local spin-density LSDA functional based on ultrasoft pseudopotentials was used to optimize the surface geometry, first, and then calculate the local density of states. A plane wave cutoff of 260eV and a (3x3x1) grid of $k$-points were chosen. We limit our discussion by presenting the calculation for an oxygen-termination on top of an Fe-octahedral layer (Fig.1(b)).

It is known that in transition metals the 3d electrons make a large contribution to the tunneling current. In the case of the oxygen-terminated surface, one may not expect a tunneling contribution from the oxygen p-states. However, on the surface the hybridization between oxygen p-states and transition metal d-states changes the situation by altering the oxygen p-orbitals from insulating to conductive[7]. The layer-projected partial density of states in Fig.3a shows significant p-d states hybridization between the topmost oxygen anions and the Fe ions in the layer underneath. The p-states of the oxygen anions are shifted towards the Fermi level eliminating the band gap. This effect almost disappears in the second oxygen layer, where the oxygen p-states are similar to those in the bulk.

As mentioned earlier, the periodicity and position of the maxima appearing around an oxygen vacancy (Fig.2(b)) correspond to iron sites in the layer below the topmost oxygen layer. As the image in Fig.2(b) was taken with a bias voltage of -1V, we should expect a significant spin-polarization at 1eV below the Fermi-level. The spin-polarization of the surface oxygen is much smaller than the spin-polarization of the Fe-states (Fig.3a). There are corresponding peaks in the majority of the Fe-octahedral d-states about 2eV and 1 eV below Fermi level, which are more likely responsible for the increase of tunneling conductivity in the vicinity of the vacancy. We therefore conclude that spin-dependent tunneling across the magnetite interface represents an interplay between oxygen and Fe contributions.

An example of how the surface states change in the presence of the oxygen vacancy is shown in Fig.3(c). The calculations were performed for a 2x2 supercell with and without a surface defect present. Fig.3b shows the majority states distribution above the surface, which displays a maximum density above each oxygen atom in the case of the defect-free surface. In the presence of the oxygen vacancy, we see that new maximas appear and the distribution is no longer regular around the vacancy (Fig.3c).

Although, the comparison of the calculations with the experiment is rather qualitative, the main conclusion can be drawn is that the spin polarization at the magnetite interface is altered locally by the presence of surface defects.

Conclusions

We have employed SP-STM to study spin transport through a magnetite interface. Significant variations in tunneling conductance are observed, suggesting that even atomic defects at an interface can have dramatic consequences for the spin-transport properties of half-metallic interfaces.

ACKNOWLEDGMENTS
This work was supported by Science Foundation of Ireland (SFI) under contract 00/PI.1/C042. N.B. thanks Ildus S. Nugmanov.

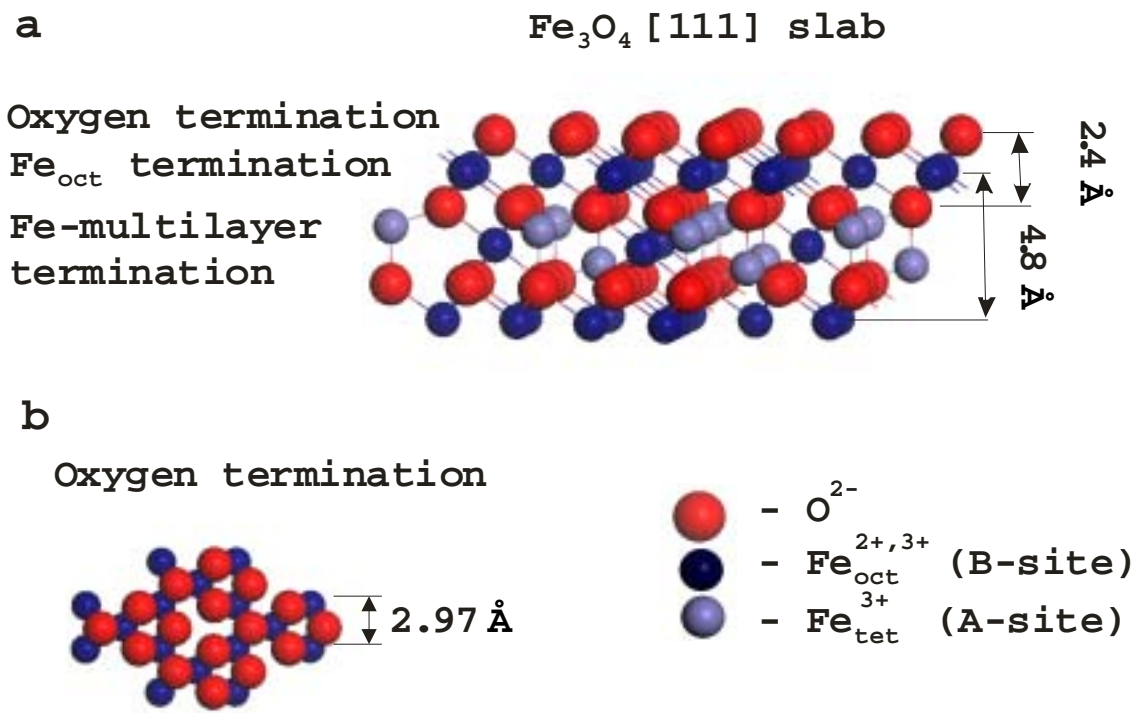

Fig.1. Magnetite [111] vacuum slab (a) and oxygen termination on top of Fe-octahedral layer (b).

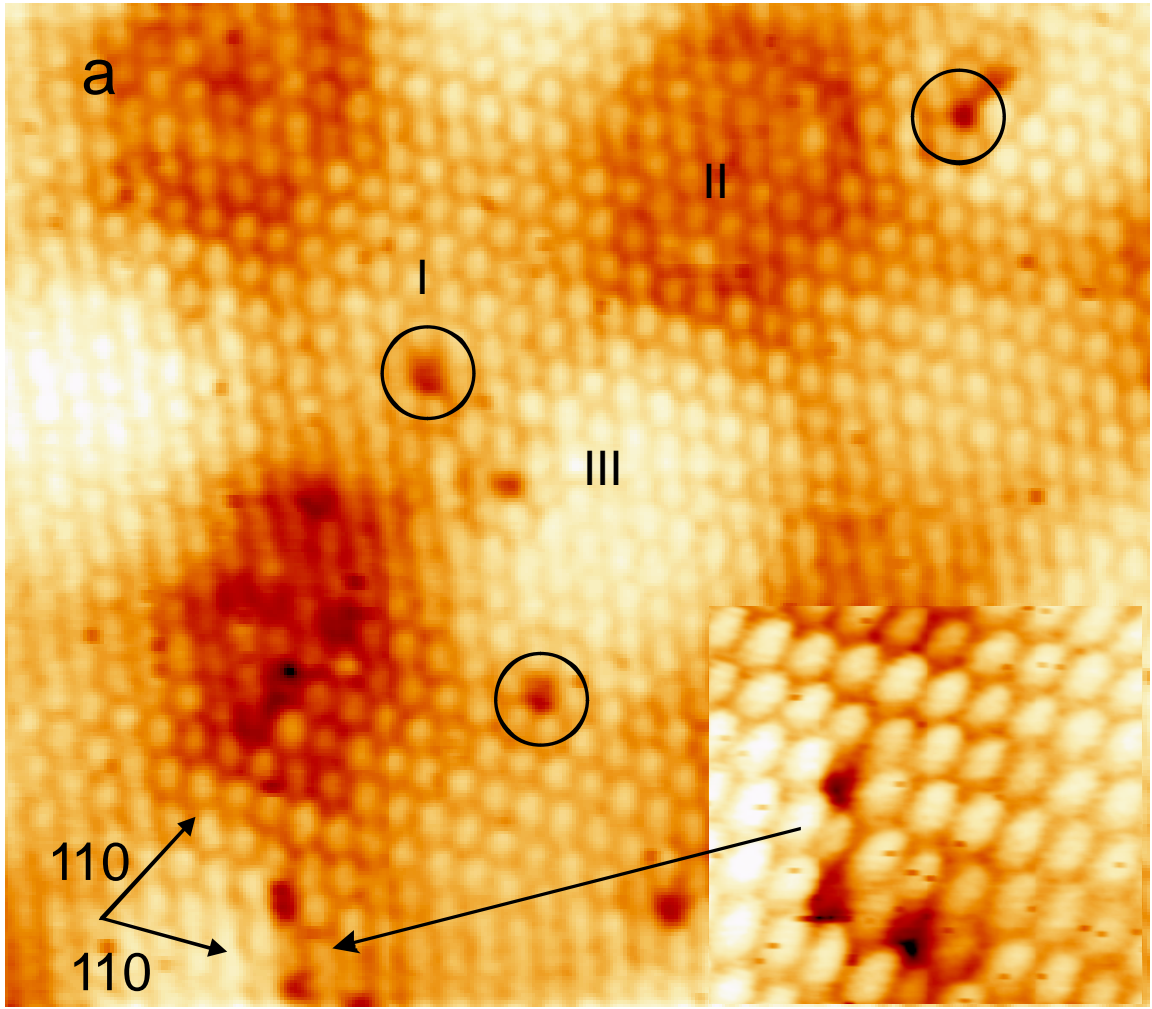

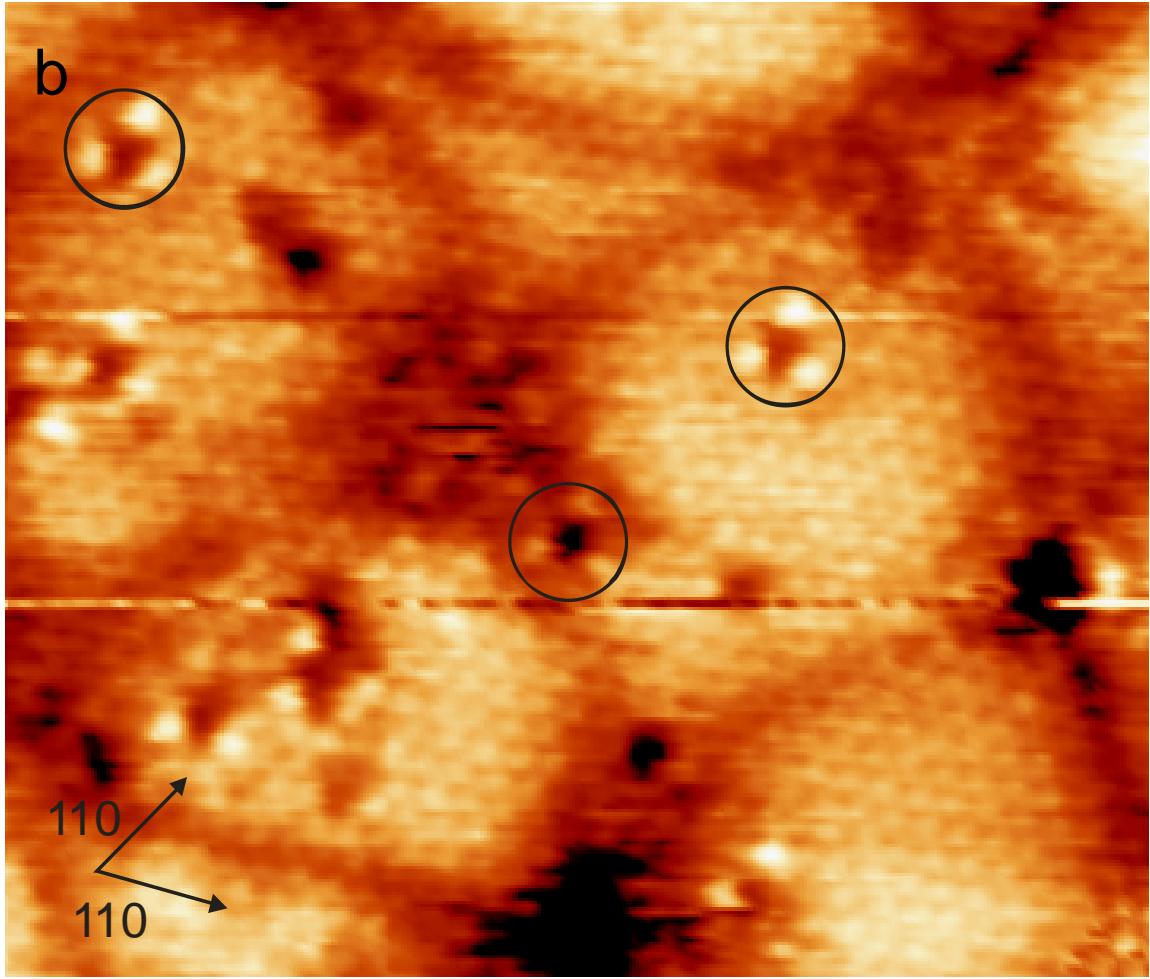

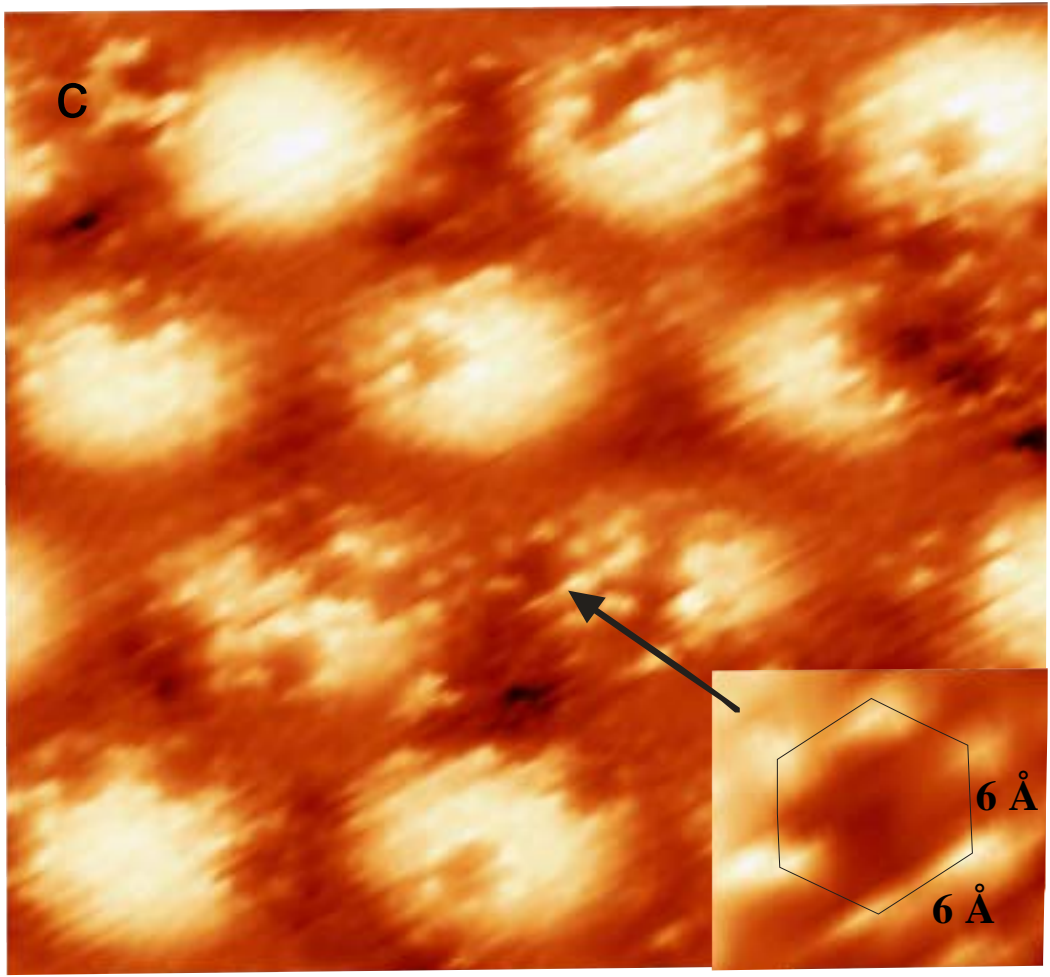

Fig.2. STM images (10.5×8.5 nm², 7×8 nm² and 12.0×17.5 nm²) of the superstructure without (a) and with magnetic field (b,c). Circles mark the oxygen vacancies in the topmost surface layer. (b) Three bright spots appeared around vacancy correspond to Fe sites with 6 Å interatomic distance; (c) A case of higher density of defects. Multiple vacancies are now surrounded by the number of bright spots, which still has the same periodicity and symmetry as in (b). ($V_{bias}$= -1.0 V, $I_t$ = 0.1 nA, MnNi tip). Inserts show a high-resolution image of the defects.

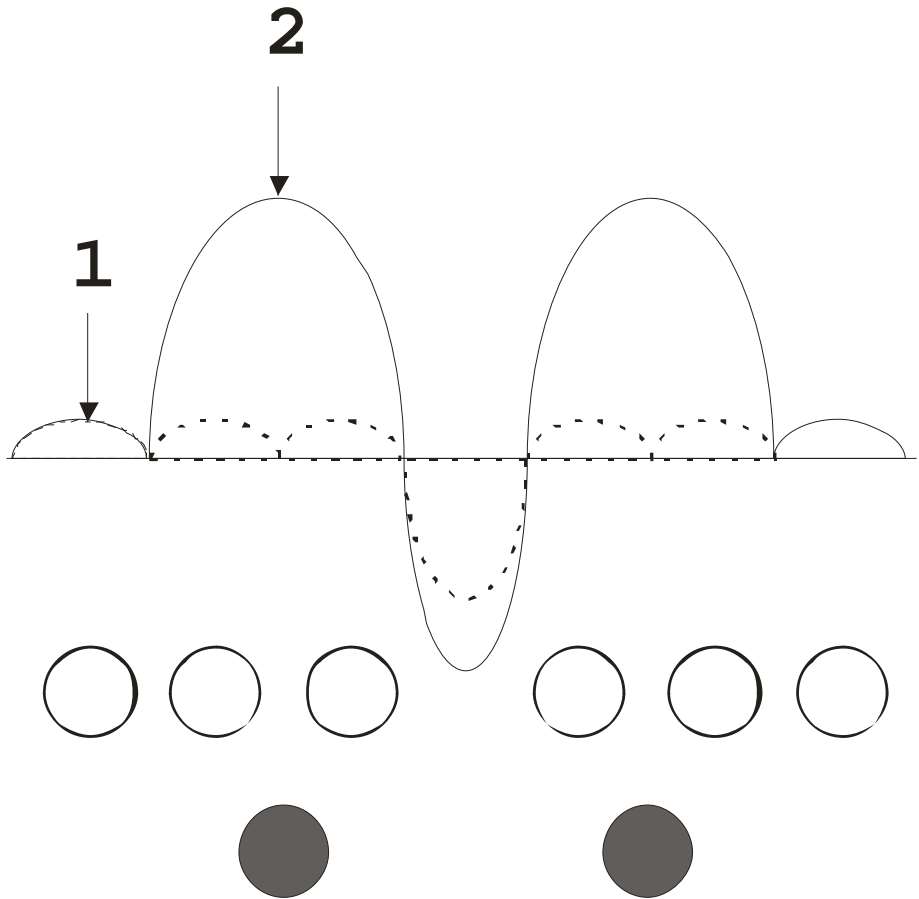

Fig.2. (d) Schematic representative of the STM line profile (solid curve – magnetic field applied; dashed line – no magnetic field) above an oxygen-defect. (Oxygen atoms are white circles, Fe atoms – grey; tip position 2 indicates the magnetic contrast maxima seen in Fig.2(b)).

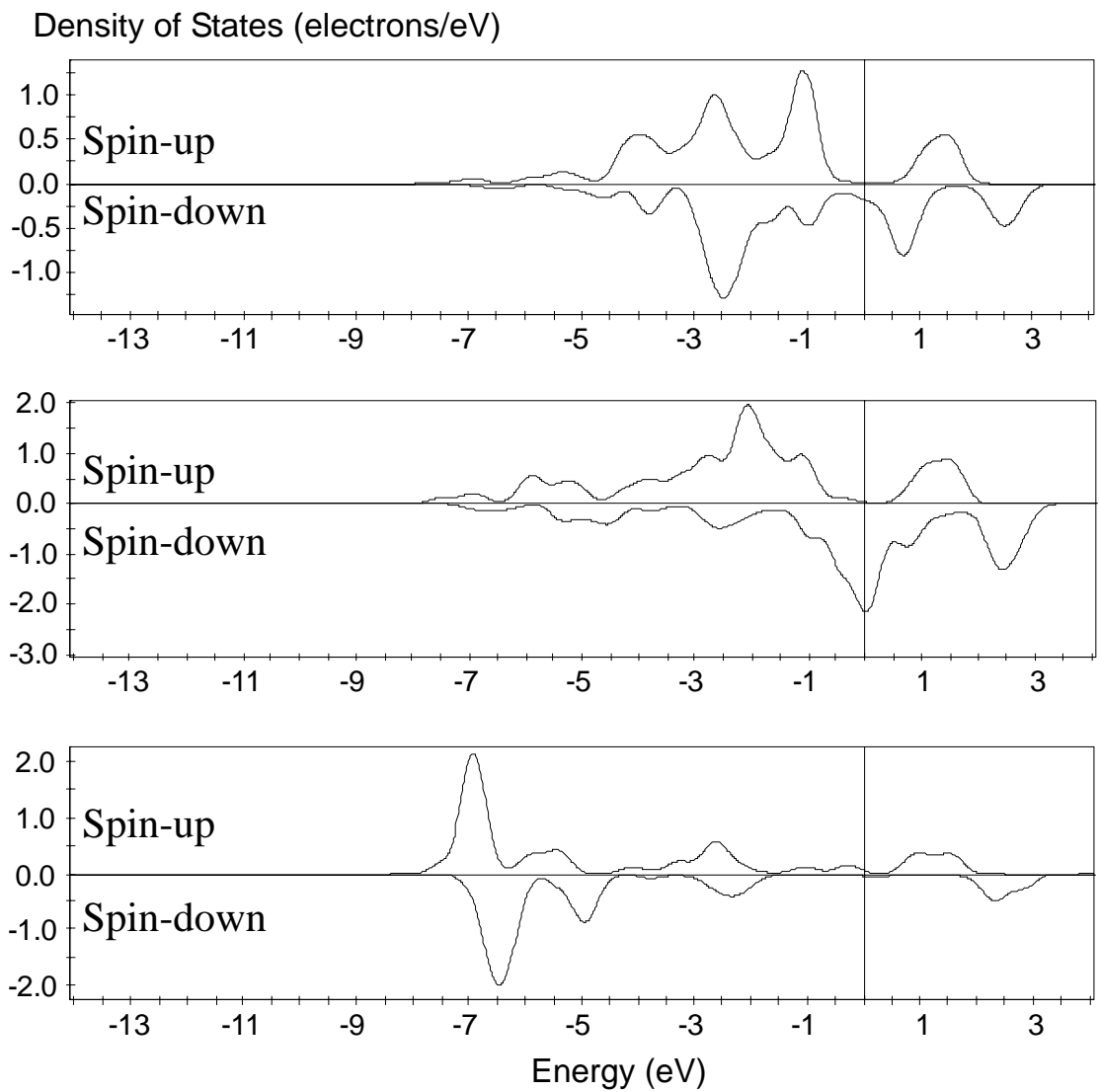

Fig. 3. (a) Partial density of states corresponding to: (top) 2p (O) states in closed-packed oxygen topmost layer; (middle) 3d (Fe) states in the Fe-octahedral layer, next to the topmost oxygen layer; (bottom) 2p (O) states in the third layer.

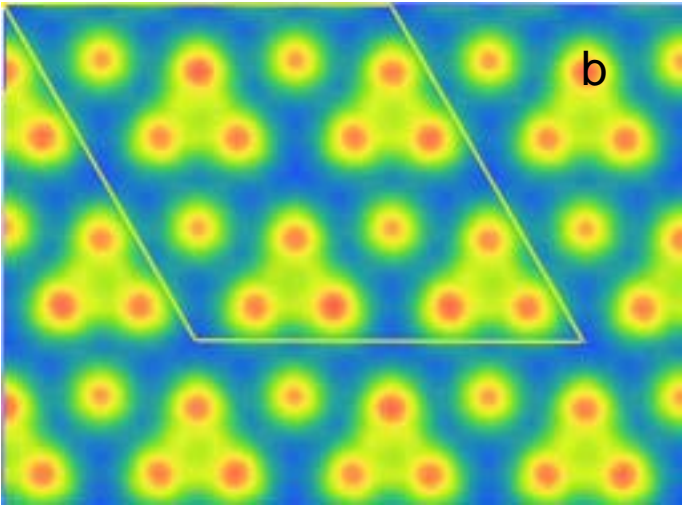

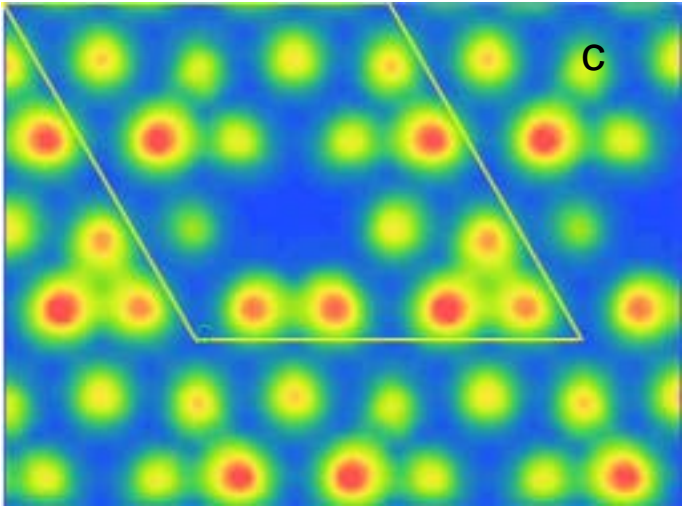

**Fig. 3. (b,c)** Spin-up electron density above the oxygen terminated surface: ideal surface layer (b); and in the presence of an oxygen vacancy (c). The sketch of energy range [$E_F$-1eV,$E_F$] was chosen to make it comparable with STM data ($V_{bias}$ = -1V). The (2x2) supercell used in the calculations is marked.

[1] G. Hu and Y. Suzuki, Phys. Rev. Lett. **89**, 276601 (2002).

[2] P. Seneor, A. Fert, J.-L. Maurice, F. Montaigne, F. Petroff, and A. Vaurès, Appl. Phys. Lett. **74**, 4017 (1999).

[3] Yu. S. Dedkov, U. Rüdiger, and G. Güntherodt, Phys. Rev. B **65**, 064417 (2002).

[4] J. S. Moodera, J. Nowak, and J. M. van de Veerdonk, Phys. Rev. Lett. **80**, 2941 (1998).

[5] H. F. Ding, W. Wulfhekel, J. Henk, P. Bruno, and J. Kirschner, Phys. Rev. Lett. **90**, 116603 (2003).

[6] S. Zhang, P. M. Levy, A. C. Marley, and S. S.P. Parkin, Phys. Rev. Lett. **79**, 3744 (1997).

[7] E. Y. Tsymbal, O. N. Mryasov, and P. R. LeClair, J. Phys.: Condens. Matter **15**, R109 (2003).

[8] I. I. Oleynik and E. Y. Tsymbal, J. Appl. Phys. **93**, 6429 Part 2 (2003).

[9] W. A. Hofer and A. J. Fisher, Surf. Sci. Lett. **515**, L487 (2002).

[10] D. Wortmann, S. Heinze, P. Kurz, G. Bihlmayer, and S. Blugel, Phys. Rev. Lett. **86**, 4132 (2001).

[11] M. Bode, Rep. Prog. Phys. **66**, 523 (2003).

[12] U. Schlickum, Appl. Phys. Lett. **83**, 2016 (2003).

[13] A. Minakov and I. V. Shvets, Surf. Sci. **236**, L377 (1990).

[14] A. Kubetzka, M. Bode, O. Pietzsch, and R. Wiesendanger, Phys. Rev. Lett. **88**, 57201 (2002).

[15] I. V. Shvets, R. Wiesendanger, D. Burgler, G. Tarrach, H. J. Guntherodt, and J.M. D. Coey, J. Appl. Phys. **71**, 5489 (1992).

[16] G. Mariotto, S. Murphy, and I. V. Shvets, Phys. Rev. B **66**, 245426 (2002).

[17] S. F. Ceballos, G. Mariotto, S. Murphy, and I. V. Shvets, Surf. Sci. **523**, 131 (2003).

[18] Sh. K. Shaikhutdinov, M. Ritter, X.-G. Wang, H. Over, and W. Weiss, Phys. Rev. B **60**, 11062 (1999).

[19] I. V. Shvets, N. Berdunov, G. Mariotto, and S. Murphy, Europhys. Lett. **63**, 867 (2003).

[20] N. Berdunov, S. Murphy, G. Mariotto, and I. V. Shvets, Phys. Rev. B **in print**, (2003); also arXiv:cond-mat/0403238

[21] W.A. Hofer, A.J. Fisher, R.A. Wolkow, P. Grutter, Pys.Rev.Lett. **87**, 236104, 2001

[22] W.A. Hofer, Prog. Surf. Sci. 71, 147, 2003

[23] M. Julliere, Phys. Lett. A **54**, 225 (1975).

[24] Segall M. D., Lindan P. J. D., Probert M. J., Pickard C. J., Hasnip P. J., Clark S. J. and Payne M. C., J. Phys. Condens. Matter, **14** (2002) 2717.